\documentclass[12pt,letterpaper]{article}
\pdfoutput=1
\usepackage[top=3cm, bottom=2cm, left=2cm, right=2cm]{geometry}
\usepackage[usenames,dvipsnames,svgnames]{xcolor}
\definecolor{darkblue}{rgb}{0.0,0.1,0.3} 
\definecolor{darkgreen}{rgb}{0,0.65,0}
\definecolor{dblue4}{rgb}{0.06,0.31,0.55} 
\definecolor{nicered}{rgb}{0.7,0.1,0.1}
\definecolor{nicegreen}{rgb}{0.1,0.5,0.1}

\usepackage[numbers,sort&compress]{natbib}
\usepackage[utf8]{inputenc}
\usepackage{amsmath,amssymb,amsfonts,amsthm}
\usepackage{xcolor}
\usepackage[colorlinks=true,citecolor= nicegreen,linkcolor=nicered]{hyperref}

\usepackage{verbatim}
\usepackage{graphicx}
\graphicspath{{figures/}}
\usepackage{cancel}
\usepackage{tipa}
\usepackage{array}   
\newcolumntype{L}{>{$}l<{$}} 
\newcolumntype{R}{>{$}r<{$}} 
\usepackage{booktabs}
\usepackage{tabularx}
\usepackage{multirow}
\usepackage{dsfont}
\newcolumntype{Y}{>{\centering\arraybackslash}X}
\usepackage{tikz}
\usepackage{slashed}
\usepackage{cancel}

\def\c{,\allowbreak}

\usepackage{comment}
\includecomment{details}
\specialcomment{details}
{\begingroup}{\endgroup}

\usepackage[affil-it]{authblk}

\newcommand{\orcid}{\includegraphics{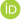}}
\newcommand{\orcidlink}[1]{\href{https://orcid.org/#1}{{\orcid}}}

\title{Dirac dark matter, neutrino masses, and dark baryogenesis.  }

\author[1,2]{Diego Restrepo\footnote{restrepo@udea.edu.co}}
\author[1]{Andrés Rivera\footnote{afelipe.rivera@udea.edu.co}}
\author[3]{Walter Tangarife\footnote{wtangarife@luc.edu}}
\affil[1]{Instituto de Física, Universidad de Antioquia,
 Calle 70 No 52-21, Medellín, Colombia}
\affil[2]{Instituto de Física Gleb Wataghin, UNICAMP, 13083-859, Campinas, SP, Brazil} 
\affil[3]{Department of Physics, Loyola University Chicago, 1032 W. Sheridan Road, Chicago, IL, 60660, USA}

\date{\small \today}

\begin{document}
\maketitle

\begin{abstract}

We present a gauged baryon number model as an example of models where all new fermions required to cancel out the anomalies help to solve phenomenological problems of the standard model (SM). Dark fermion doublets, along with the iso-singlet charged fermions, in conjunction with a set of SM-singlet fermions, participate in the generation of small neutrino masses through the Dirac-dark Zee mechanism. The other SM-singlets explain the dark matter in the Universe, while their coupling to an inert singlet scalar is the source of the $CP$ violation. In the presence of a strong first-order electroweak phase transition, this ``dark'' $CP$ violation allows for a successful electroweak baryogenesis mechanism. 
\end{abstract}

\section{Introduction}
In this work, we present a self-contained framework to explain the main phenomenological problems of the standard model (SM) by using a dark sector with a gauged baryon or lepton number Abelian symmetry, both of which allows for a wide range of $Z'$ masses and gauge couplings~\cite{Carone:1994aa,Carone:1995pu}. 
In this kind of models, fermionic dark matter serves as the source of the $CP$ violation required to generate the baryon asymmetry of the Universe~\cite{Carena:2018cjh,Carena:2019xrr}. 
We also search for all the solutions to the anomaly conditions 
compatible with the effective Dirac-neutrino mass operator, focusing on their scotogenic realizations. In this setup, we can have a dark sector that contains proper dark matter candidates and sources both the baryon asymmetry of the Universe and the smallness of neutrino masses.

The minimal field content for a gauged baryon or lepton number requires the inclusion 
of an electroweak-scale set of fields consisting of a vector-like fermion doublet and a vector-like iso-singlet~\cite{FileviezPerez:2010gw,FileviezPerez:2011pt,Ma:2020quj}, which behave as \emph{anomalons} of the Abelian symmetry~\cite{Carena:2018cjh,Carena:2019xrr}. The corresponding chiral
fields acquire masses from the Abelian gauge symmetry's spontaneous symmetry breaking (SSB) 
at some higher energy scale. This condition fixes the absolute value of the baryon or lepton number charge of the complex SM-singlet scalar responsible for the SSB. 

Extra SM-singlets are required to cancel out the anomalies. In this work, we show that the anomaly conditions for a gauged baryon or lepton number symmetry can be expressed in terms of the well known set of Diophantine equations
\begin{align}
    \sum_{\alpha=1}^N z_\alpha=&0\,,&
    \sum_{\alpha=1}^N z_\alpha^3=&0\,,
\end{align}
where $z_\alpha$ can be chosen as integers without loss of generality. The data set with the full set of solutions with $N\le 12$ and $|z_\alpha|\le 30$~\cite{Restrepo:2021kpq}, can be used directly here. From there, we can choose the solutions with at least two repeated charges
to be assigned to the right-handed neutrinos in the effective Dirac-neutrino mass operator with the complex SM-singlet scalar to some nonzero power.

We further filter out solutions for which the extra SM-singlet fermions can get masses from the SSB of the Abelian symmetry. We found 951 $\operatorname{U}(1)_B$ solutions that are classified into 25 types of multi-component~\cite{Boehm:2003ha} and 
multi-flavor~\cite{Herms:2021fql} dark matter models. We focus on the phenomenology of the minimal $\operatorname{U}(1)_B$ solution, which allows for effective Dirac-neutrino masses, and contains just a Dirac-fermionic dark matter (DM) particle.
The DM fields play a leading role as the source of the $CP$ violation required for generating the baryon asymmetry~\cite{Carena:2018cjh} through the Yukawa coupling with one extra SM-singlet scalar (with the same Abelian charge as the scalar responsible for the SSB). The potential for this scalar and the Higgs leads to the strong first-order electroweak phase transition (EWPT) required to have successful electroweak baryogenesis~\cite{Cline:2017qpe}.
This SM-singlet scalar can acquire a vacuum expectation value (VEV) at high temperatures, increasing the tree-level potential barrier and allowing for a stronger EWPT.

The dark $CP$ violation is diffused in front of the bubble walls, creating a chiral asymmetry for the DM particle. 
The asymmetry is transferred to the SM sector by the time-like component of the $Z'$ background, which serves as the baryon number chemical potential for the SM quarks through the anomalous currents of the $Z'$~\cite{Carena:2018cjh,Carena:2019xrr}, resulting in a net baryon asymmetry inside the bubble. 
Since $CP$ is spontaneously violated in the dark sector, the model is safe against contributions to
quark electric dipole moments.

The required heavy anomalons can also be used to realize the effective Dirac-neutrino mass operator through the Dirac-Dark Zee radiative seesaw mechanism. In this way, the dark sector can also explain the smallness of the neutrino masses. As in the original Dirac-Zee model, the extra scalar content is just a set of iso-singlet charged scalars~\cite{Nasri:2001ax}, which does not affect the anomaly cancellation conditions.

Since the right-handed neutrinos can be thermalized in the early universe through their $Z'$ interactions with the SM quarks, we need to remain consistent with cosmological constraints on the effective number of relativistic
degrees of freedom $N_{\text{eff}}$~\cite{Dolgov:2002wy}. This constraint goes in the same direction as the constraints from 
direct detection cross section of the Dirac-fermionic dark matter with the nucleon, requiring large vacuum expectation values for
the scalar field responsible for the SSB of $\operatorname{U}(1)_B$. However, we still have room to explain the SM's three
more critical phenomenological problems, which can be fully explored in future experiments.

The structure of this article is the following, in Section~\ref{sec:anomaly}, we present the general framework to define a gauged baryon o lepton number Abelian symmetry. In Section~\ref{sec:U1B-model}, we focus on the minimal $\operatorname{U}(1)_B$ model with a Dirac-dark Zee mechanism to generate small neutrino masses and a Dirac-fermionic dark matter candidate. In Section~\ref{sec:baryogenesis}, we establish the conditions to have successful electroweak baryogenesis and perform the random scan to obtain the models which satisfy all the experimental constraints, including the proper $\Delta N_{\text{eff}}$. Finally, we present our conclusions in Section~\ref{sec:con}.

\section{Local $\operatorname{U}(1)_X$ extension of the standard model}
\label{sec:anomaly}

We extend the standard model (SM) by adding a general local $\operatorname{U}(1)_X$ symmetry and a set of new chiral fields with $X$-charges as displayed in Table~\ref{tab:prtcntnt}, including $N'$ SM-singlet chiral fermions, two iso-singlet charged fermions $e'_R$ and $e''_L$, and two $\operatorname{SU}(2)_L$ doublets with components 
$
    L'_L=\begin{pmatrix}
N'_L & e'_L
\end{pmatrix}^{\operatorname{T}}
$ and
$
L''_R=\begin{pmatrix}
N''_R & e''_R
\end{pmatrix}^{\operatorname{T}}
$. It is worth noting that, without this
minimal set of extra chiral fields with non-zero hypercharge, it is not possible to have zero lepton (quark) $X$-charges, as required by a gauged baryon (lepton) number symmetry~\cite{FileviezPerez:2010gw,Ma:2020quj}.

\begin{table}
  \centering
  \begin{tabular}{l|c|c|c}\hline
    Field&$\operatorname{SU}(2)_L$ & $\operatorname{U}(1)_Y$&$\operatorname{U}(1)_X$\\\hline
     $u_{Ri}$      & $\mathbf{1}$ & $2/3$   & $u$   \\ 
     $d_{Ri}$      & $\mathbf{1}$ & $-1/3$   & $d$   \\ 
     $\left(Q_i\right)^\dagger$      & $\mathbf{2}$ & $-1/6$   & $Q$   \\ 
     $\left(L_i\right)^\dagger$      & $\mathbf{2}$ & $1/2$   & $L$   \\ 
     $e_{R_i}$      & $\mathbf{1}$ & $-1$   & $e$   \\ 
     $\left(L'_L\right)^\dagger$      & $\mathbf{2}$ & $1/2$   & $-x'$   \\ 
     $e'_R$      & $\mathbf{1}$ & $-1$   & $x'$   \\ 
     $L''_R$      & $\mathbf{2}$ & $-1/2$   & $x''$   \\ 
     $\left(e''_L\right)^\dagger$      & $\mathbf{1}$ & $1$   & $-x''$   \\ 
     $\chi_\alpha$      & $\mathbf{1}$ & $0$   & $z_\alpha$   \\\hline
  \end{tabular}
  \caption{Fermion content and its quantum numbers. $i=1,2,3$, $\alpha=1,2,\ldots,N'$.}
  \label{tab:prtcntnt}
\end{table}

There are several conditions that we need to impose on the charge assignment to achieve the goals of our model:
\begin{enumerate}
    \item The model must be anomaly free.
    \item At least two of the new SM-singlet chiral fields $\chi_\alpha$ must correspond to the right-handed neutrinos associated with light Dirac neutrino masses. We require that these fields have the same $\operatorname{U}(1)_{X}$ charge.
    \item Two SM-singlet chiral fields, say $\chi_R$ and $\chi_L$, can form the Dirac-DM particle that also participates in the baryon asymmetry generation. These fields couple to a heavy scalar field $\Phi$ that spontaneously breaks $\operatorname{U}(1)_{X}$ and provides mass to the chiral fields. For simplicity, we assume that the same scalar field provides the mass for the heavy doublets $L'_L$ and $L''_R$ (and the heavy iso-singlets $e'_R$ and $e''_L$). These requirements impose the following conditions on the $X$-charges
\begin{subequations}    
    \begin{align}
    \label{eq:chirmass1}
        q_\Phi = & q_{\chi_R}+q_{(\chi_L)^\dagger}\, ,\\
    \label{eq:chirmass2}
        q_\Phi = & q_{(L'_L)^\dagger} + q_{L''_R}= -\left[q_{e'_R}+q_{(e''_L)^\dagger}\right]= -x'+x''\,.
    \end{align}
\label{eq:chirmass}
\end{subequations}    
\end{enumerate}

The anomaly cancellation conditions on $\left[\operatorname{SU}(3)_c\right]^{2}  \operatorname{U}(1)_{X}$, $\left[\operatorname{SU}(2)_{L}\right]^{2} \operatorname{U}(1)_{X}$, $\left[\operatorname{U}(1)_{Y}\right]^{2} \operatorname{U}(1)_{X}$,  allow us to express three of the $X$-charges in terms of the others
\begin{align}\label{eq:sol0}
  u=&-e-\frac23 L-\frac19\left(x'-x''\right)\,,& 
  d=& e+\frac43 L-\frac19\left(x'-x''\right)\,,& 
  Q=& -\frac13 L+\frac19\left(x'-x''\right)\,,
\end{align}
 while the $\left[\operatorname{U}(1)_{X}\right]^{2} \operatorname{U}(1)_{Y}$ anomaly condition reduces to
 \begin{align}
     (e+L)(x'-x'')=0\,.
 \end{align}

Note that the vector-like solution $x''=x'$ leads to the same solution as the SM extension with only extra singlet chiral fermions with no hypercharge in Ref.~\cite{Bernal:2021ezl}. This kind of solution is incompatible with a gauged baryon or lepton number and will not be considered here. To cancel out the $\left[\operatorname{U}(1)_{X}\right]^{2} \operatorname{U}(1)_{Y}$ anomaly, we choose instead
\begin{align}
\label{eq:emL}
    e=-L\,,
\end{align}
so that~\cite{Carena:2019xrr,Ma:2021fre}
\begin{align}
\label{eq:Q}
    Q=-u=-d=-\frac13 L+\frac19\left(x'-x''\right).
\end{align}
Notice that, because of Eqs.~\eqref{eq:emL} and \eqref{eq:Q}, the $X$-charge of the Higgs must always be zero in these scenarios.

The gravitational anomaly, $\left[\operatorname{SO}(1,3)\right]^2\operatorname{U}(1)_Y$, and the cubic anomaly, $\left[\operatorname{U}(1)_X\right]^3$,
can be written as the following system of Diophantine equations, respectively,
\begin{align}
    \label{eq:Dcoond}
    \sum_{\alpha=1}^N z_\alpha=&0\,,&
    \sum_{\alpha=1}^N z_\alpha^3=&0\,,
\end{align}
where $N=N'+5$ and
\begin{align}
\label{eq:Nalpha}
    z_{N'+1}=&-x'\,, &&& z_{N'+2}=&x''\,,\nonumber\\
   & &z_{N'+2+i}&=L\,,\quad i=1,2,3\,,&&
\end{align}
It worth noticing that, to our knowledge, it is the first time that the gravitational and cubic anomaly equations are expressed in the most general way for the gauge baryon or lepton number Abelian symmetries. This will allows us to use the already known general solutions to the Diophantine equations~\eqref{eq:Dcoond} for local $\operatorname{U}(1)$ symmetries with only extra SM-singlet chiral fermions~\cite{Costa:2019zzy}, but assigned to the new fields.  In fact, until now, only specific solutions to the full set of anomaly conditions for the gauge baryon or lepton number Abelian symmetries have been reported in the literature so far\footnote{The vector-like solution for $\operatorname{U}(1)_L$ and $\operatorname{U}(1)_B$ in~\cite{Carena:2019xrr}, in the proposed ordering, are respectively $(1\c 1\c 1\c       q\c -q-N_g     \c -q\c q+N_g\c -1\c -1\c -1)$ and $(q\c -N_g-q\c -q\c N_g+q)$ with massless neutrinos in the last case. While the chiral $\operatorname{U}(1)_L$ solution in~\cite{Ma:2021fre} is $(2\c  2\c  2\c  -3\c   6\c -4\c  -5\c -3\c 0\c  1\c  1\c  1)$. }.

Notice that any solution of Eqs.~\eqref{eq:Dcoond} can be readily interpreted as gauged baryon number symmetry, $\operatorname{U}(1)_B$, if we do not assign any integer of the solution to $L$, so that it remains zero. 
The simplest solution to have $\operatorname{U}(1)_B$ is for $N'=2$  with a massive SM-singlet Dirac fermion. Then, in addition to fixing $L=0$, we can choose $z_3=-z_1=-x'$ and $z_4=-z_2=x''$~\cite{Ma:2020quj}. However, in this kind of solution, neutrinos are still massless, as in the SM. We are interested here in solutions where the right-handed neutrinos are also charged under $\operatorname{U}(1)_B$ or $\operatorname{U}(1)_L$

To have a gauge lepton number symmetry $\operatorname{U}(1)_L$, however, we require two additional conditions on the solution of of Eqs.~\eqref{eq:Dcoond}:  (a) the set must have three repeated integers whose value must be assigned to $L$; and, since Eq.~\eqref{eq:Q} can be rewritten as
\begin{align}
\label{eq:Qzero}
    9\,Q=-\sum_{\alpha=N'+1}^{N'+5}z_\alpha\,,
\end{align}
(b) the corresponding subset of integers $z_\alpha$ in \eqref{eq:Nalpha} must add to zero. 

In this way, the previously found solutions for SM-singlet chiral extensions of the SM with effective Dirac-neutrino masses~\cite{Restrepo:2021kpq} can be used directly here.
There, a SM-singlet scalar $\Phi$ appears in the effective Dirac-neutrino mass operator~\cite{Yao:2018ekp}, in terms of Weyl fermions,
\begin{align} 
\label{eq:nmo56}
    \mathcal{L}_{\text{eff}} = h_{\nu}^{\alpha i} \, \left( \nu_{R\alpha}\right)^{\dagger} \, \epsilon_{ab} \, L_i^a \, H^b \left(\frac{\Phi^*}{\Lambda}\right)^\delta + \text{H.c.},\qquad \text{with $i=1,2,3$}\,,
\end{align}
and $\delta = 1,2,\ldots$ for dimension-$d=4+\delta$ operators. We only consider solutions with at least two repeated charges, $\nu$, to be assigned to the right-handed neutrinos, $\nu_{R\alpha}$,
\begin{align}
    \chi_1\to \nu_{R1},\cdots, \chi_{N_\nu}\to \nu_{R\, N_\nu},\qquad 2\le N_\nu\le 3\,,
\end{align}
and with
the scalar $\Phi$ providing mass to all pairs of SM-singlet chiral fields, like the ones in Eq.~\eqref{eq:chirmass1}. Since all the solutions in~\cite{Restrepo:2021kpq}
have at least one massive SM-singlet Dirac fermion,  it is always possible to reassign it to $x'$ and $x''$ as in Eq.~\eqref{eq:chirmass2}, so that all the solutions in~\cite{Restrepo:2021kpq} satisfy
\begin{align}
\label{eq:phln}
q_\Phi=&\frac{L-\nu}{\delta}\,,&   |q_\Phi|=&|-x'+x''|,&
|q_\Phi|=&|\chi_\alpha+\chi_\beta|\qquad \alpha,\beta=N_\nu,\ldots,N'\,.
\end{align}
Then, we can reinterpret each one of the dark symmetries, $\operatorname{U}(1)_D$, in~\cite{Restrepo:2021kpq} as a gauged baryon number symmetry, $\operatorname{U}(1)_B$\footnote{The type of solutions with $m=0$ in Table~1 of \cite{Restrepo:2021kpq}.}.
Note, however, that the integers which solve the two Diophantine equations~\eqref{eq:Dcoond} in both cases, are now assigned to different fields. While in the dark symmetry case the integers of the solutions corresponds to charges of extra SM-singlet chiral fermion fields, and the SM charges are neutral under the new gauge symmetry; in the gauged baryon number model, two of the integers need to be assigned to the set of non-zero hypercharge fermion doublets and iso-singlets, and only the SM-lepton sector is neutral under the new symmetry. This leads to completely new phenomenology with contributions to $N_{\text{eff}}$, flavor observables and direct detection and collider constraints on $Z'$, which will be studied below for one specific solution.
On the other hand, the active symmetry solutions $\operatorname{U}(1)_X$ in~\cite{Restrepo:2021kpq}, which also require three repeated charges,  need to be checked against the extra conditions from Eq.~\eqref{eq:chirmass2} and Eq.~\eqref{eq:Qzero}, with $Q=0$, before they can be identified as a gauged lepton number symmetry, $\operatorname{U}(1)_L$.

\begin{figure}
    \centering
    \includegraphics[scale=0.65]{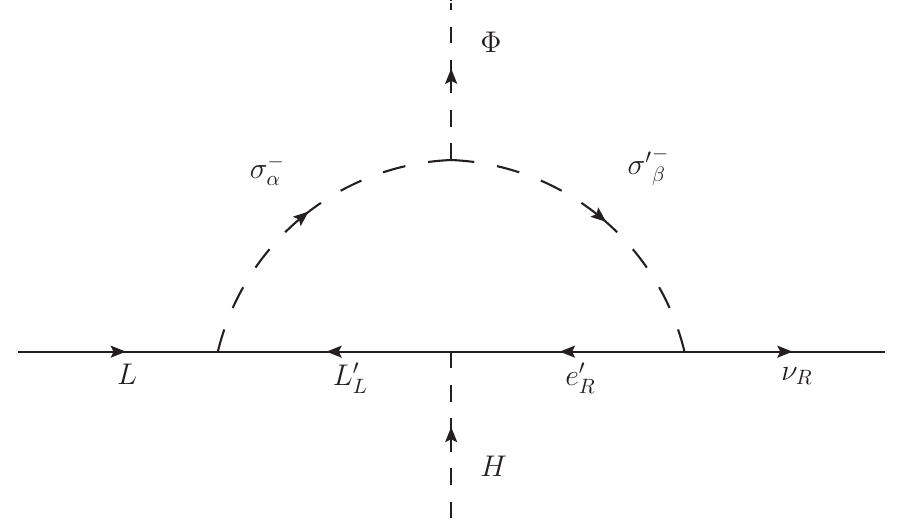}
    \caption{Diagram for the Dirac-dark Zee model.}
    \label{fig:zee}
\end{figure}

We are interested in the case of a scotogenic scenario where the dark sector participates in the radiative neutrino mass loop. To realize the effective Dirac-neutrino mass operator with $d=5$,
we consider the Dirac-Dark Zee topology as shown in the diagram in Fig.~\ref{fig:zee}. In order to have a rank-2 light Dirac-neutrino mass matrix, we require the addition of two sets of two iso-singlet charged scalars $\sigma^-_\alpha$ and ${\sigma'}^-_\alpha$, ($\alpha=1,2$) with $X$-charges $\sigma$ and $\sigma'$ respectively.
Then, we have 
\begin{align}
\label{eq:cndtns}
    {\sigma}=&L+x'\,,& {\sigma'}=&\nu+x'\,,& q_\Phi=&\sigma-\sigma'\,,
\end{align}
and we can always recover the conditions in Eq.~\eqref{eq:phln} for $\delta=1$. In this way, this topology can be realized for all the $\delta=1$ solutions in~\cite{Restrepo:2021kpq}. 
The same happens for any other topology that realizes the effective Dirac-neutrino mass operator since the addition of the scalars does not affect the anomaly cancellation. In fact, once the anomaly conditions are satisfied, any extra fields that are required to allow for the mass operator must be either scalar of vector-like fermions.

\section{An explicit implementation for $\operatorname{U}(1)_B$}
\label{sec:U1B-model}

In this work, we consider a specific solution to the conditions from Section~\ref{sec:anomaly} in order to realize neutrino masses, dark matter, and baryogenesis. 
 We analyze the integer set $(5, 5,-2, -3, 1, -6)$ in Ref.~\cite{Restrepo:2021kpq} (ordered according to Eq.~\eqref{eq:Nalpha}), which is a solution to Eq.~\eqref{eq:Dcoond} and allows the effective Dirac-neutrino mass operator at $d=5$. We can reinterpret this as a local $\operatorname{U}(1)_B$ with $Q=5/9$, $x'=-1$, $x''=-6$ and the particle content shown in Table~\ref{tab:implementationU1B},
up to a global factor of $-3/5$. 
This factor has been included in the
column $\operatorname{U}(1)_B$ of Table~\ref{tab:implementationU1B}. 

\begin{table}
  \centering
  \begin{tabular}{l|c|c|c}\hline
    Field&$SU(2)_L$ & $U(1)_Y$&$\operatorname{U}(1)_B$\\\hline
     $u_{Ri}$      & $\mathbf{1}$ & $2/3$   & $u=1/3$   \\ 
     $d_{Ri}$      & $\mathbf{1}$ & $-1/3$   & $d=1/3$   \\ 
     $\left(Q_i\right)^\dagger$      & $\mathbf{2}$ & $-1/6$   & $Q=-1/3$   \\ 
     $\left(L_i\right)^\dagger$      & $\mathbf{2}$ & $1/2$   & $L=0$   \\ 
     $e_R$      & $\mathbf{1}$ & $-1$   & $e=0$   \\ 
     $\left(L'_L\right)^\dagger$      & $\mathbf{2}$ & $1/2$   & $-x'=-3/5$   \\ 
     $e'_R$      & $\mathbf{1}$ & $-1$   & $x'=3/5$   \\ 
     $L''_R$      & $\mathbf{2}$ & $-1/2$   & $x''=18/5$   \\ 
     $\left(e''_L\right)^\dagger$      & $\mathbf{1}$ & $1$   & $-x''=-18/5$   \\ 
     $\nu_{R,1}$   & $\mathbf{1}$ & $0$   & $-3$   \\
     $\nu_{R,2}$   & $\mathbf{1}$ & $0$   & $-3$   \\
     $\chi_{R}$      & $\mathbf{1}$ & $0$   & $6/5$   \\
     $\left(\chi_{L}\right)^\dagger$      & $\mathbf{1}$ & $0$   & $9/5$   \\ \hline
     $H$      & $\mathbf{2}$ & $1/2$   & $0$   \\
     $S$      & $\mathbf{1}$ & $0$   & $3$   \\
     $\Phi$      & $\mathbf{1}$ & $0$   & $3$   \\
     ${\sigma}^-_\alpha$      & $\mathbf{1}$ & $1$   & $3/5$   \\
     ${\sigma'}^-_\alpha$      & $\mathbf{1}$ & $-1$   & $-12/5$   \\
     \hline
  \end{tabular}
  \caption{Fermion (top) and scalar (bottom) content and its quantum numbers, $i=1,2,3$, $\alpha=1,2$ with proper normalized baryon number charges with a global factor $-5/3$.}
  \label{tab:implementationU1B}
\end{table}

With the new $\operatorname{U}(1)_Y$ charged chiral fermions and the two sets of charged iso-singlet scalars (${\sigma}^+_\alpha$, ${\sigma'}^-_\alpha$), it is now possible to realize the scotogenic Dirac-Dark Zee topology as displayed in the diagram in Fig.~\ref{fig:zee}.
The new charge assignment allows for the following terms in the Lagrangian 
\begin{align}
\label{eq:lagrangian-U1B}
-\mathcal{L} & \supset  h_{a,\Phi} \left(\chi_L\right)^\dagger \chi_R \Phi^* +  h_{b,\Phi} \left(e'_R\right)^\dagger e''_L \Phi^* + h_{c,\Phi} (L'_L)^\dagger L''_R \Phi^* + h^{i\alpha}_d L'_L L_i {\sigma}^+_{\alpha}\nonumber \\
  & + h_{a,S} \left(\chi_L\right)^\dagger \chi_R S^* +  h_{b,S} \left(e'_R\right)^\dagger e''_L S^* + h_{c,S} (L'_L)^\dagger L''_R S^*\nonumber \\
  & + h_e^{\alpha \beta} \nu_{R,\alpha}  e'_R {\sigma'}^+_{\beta} +  h_g H^\dagger \left(e'_R\right)^\dagger L'_L  +h_h (L''_R)^\dagger e''_L H + {\rm \,h.c.}\,,
\end{align}
where $\alpha,\beta=1,\,2$ and $i=1,\,2,\,3$. The $h$'s are Yukawa couplings, which we assume to be real parameters for the sake of simplicity, except the parameter $h_{a,S}$ which remains complex and will lead to $CP$ violation in the model as in the baryogenesis mechanism studied in Refs.~\cite{Carena:2019xrr,Carena:2018cjh}. We will further explore this mechanism in Sec.~\ref{sec:baryogenesis}. Finally, the Lagrangian also contains the scalar potential
\begin{align}
\label{eq:scalar-potential}
V(H,S,\Phi,{\sigma}^\pm_\alpha,{\sigma'}^\pm_\alpha) =&V(H)+V(S)+V(\Phi)+V({\sigma}^\pm_\alpha)+V({\sigma'}^\pm_\alpha) \nonumber\\
    &+\left[\kappa^{\alpha \beta}_S S {\sigma}^+_{\alpha} {\sigma'}^-_{\beta}
    +\kappa_{\Phi}^{\alpha\beta} \Phi {\sigma}^+_{\alpha} {\sigma'}^-_{\beta}\ 
    +\lambda'_{S\Phi}\left(S^*\Phi\right)^2
    +\text{h.c.}
    \right]\,,
\end{align}
where
\begin{align}
    V(\phi_i)=&\mu_{\phi_i}^2 \phi_i^\dagger\phi_i
    +\lambda_{\phi_i} \left(\phi_i^\dagger\phi_i\right)^2
    +\lambda_{\phi_i\phi_j}\phi_i^\dagger\phi_i\phi_j^\dagger\phi_j
    +\left(\lambda_{\phi_i S\Phi}\phi_i^\dagger\phi_i S^*\Phi+\text{h.c.}\right)\,,
\end{align}
with $\phi_i=H,S,\Phi,{\sigma}^+_\alpha,{\sigma'}^-_\alpha$, and $i<j$\,. 

\subsection{Symmetry breaking and spectrum}
\label{sec:spectrum}

In this model, the scalar $\Phi$ develops a vacuum expectation value (VEV), $\langle \Phi\rangle=v_{\Phi}/\sqrt{2}$ that remains constant as the Universe evolves near the EWPT, where the SM Higgs develops its VEV, $\langle H\rangle=v/\sqrt{2}$, with $v=246.2$ GeV.
At zero temperature, the scalar fields $S,\sigma_\alpha^{+},\sigma'^{-}_\alpha$  do not obtain a VEV.

At zero temperature, after the electroweak symmetry breaking, the scalars $H=\begin{pmatrix}G^+ & H^0\end{pmatrix}^{\operatorname{T}}$ and $\Phi$ are mixed.
In the basis $(h^0,\Phi^0)$, the mass mixing matrix is given by
\begin{align}
\label{eq:neutral-higgs-mixing-matrix}
m_h^2=\left(
\begin{array}{cc}
 -3 \lambda _H v^2+\mu _H^2-\frac{1}{2} v_{\Phi }^2 \lambda _{H \Phi } & -v v_{\Phi } \lambda _{H \Phi } \\
 -v v_{\Phi } \lambda _{H \Phi } & -\frac{1}{2} \lambda _{H \Phi } v^2+\mu _{\Phi }^2-3 v_{\Phi }^2 \lambda _{\Phi } \\
\end{array}
\right)\,,
\end{align}
where $H^0=(h^0+v+i\, G^0)/\sqrt{2}$, $\Phi=(\Phi^0+v_\Phi+i\, G'^0)/\sqrt{2}$ and $G^\pm$, $G^0$, $G'^0$ are Goldstone bosons.
This matrix is diagonalized by a unitary transformation 
\begin{align}
Z^h m_h^2 Z^{h\dagger}=m_{h,\text{diag}}^2\,,
\end{align}
such that
\begin{align}
\label{eq:alpha-mixing-matrix}
    \begin{pmatrix}
    h^0\\
   \Phi^0\\
     \end{pmatrix}=Z^h
         \begin{pmatrix}
    h_1\\
    h_2\\
     \end{pmatrix}=\begin{pmatrix}
    \cos\alpha & \sin\alpha\\
    -\sin\alpha & \cos\alpha\\
     \end{pmatrix}\begin{pmatrix}
    h_1\\
    h_2\\
     \end{pmatrix}\,.
\end{align}

The fields ${\sigma}^+_\alpha$ and ${\sigma'}^-_\alpha$ mix among themselves and generate four heavy charged scalars that will play an important role in the generation of the neutrino masses via one-loop as displayed in the Feynman diagram in Fig.~\ref{fig:zee}. In the basis $({\sigma}^\pm_\alpha,{\sigma'}^\pm_\beta)$, the charged-scalar mixing matrix is given by
\begin{align}
\label{eq:charge-higgs-mixing-matrix}
m_H^{\pm 2}=\left(
\begin{array}{cc}
 \dfrac{1}{2}\lambda_{H,{\sigma}^\pm_{\alpha}}v^2 + \mu_{{\sigma}^+_\alpha}^2 & \dfrac{1}{\sqrt{2}}\kappa_{\Phi}^{\alpha\beta} v_\Phi \\
 \dfrac{1}{\sqrt{2}}(\kappa_{\Phi}^{\alpha\beta})^T v_\Phi &  \dfrac{1}{2}\lambda_{H,{\sigma'}^\pm_{\alpha}}v^2 + \mu_{{\sigma'}_\alpha}^2 \\
\end{array}
\right)\,,
\end{align}
which is diagonalized by a $Z^+$ matrix, such that
\begin{align}
Z^+m_H^{2\pm }Z^{+\dagger}=m^2_{H^{\pm},\text{diag}}\,,
\end{align}
where
\begin{align}
\label{eq:charge-higgs-diagonalization}
\begin{pmatrix}
{\sigma}^+_1\\
{\sigma}^+_2\\
{\sigma'}^+_1\\
{\sigma'}^+_2
\end{pmatrix}
=\sum_{i,j=1}^4Z^+_{ji}H_j^+\,.
\end{align}
Also, there is a heavy scalar $S$ that will play an important role in the baryogenesis mechanism.

Regarding the new charged fermions shown in Table~\eqref{tab:implementationU1B}, we have that, in the basis $E_{Li}=(e'_L\,, e^{''}_L)$, $E_{Ri}^{\dagger}=(e^{'}_R\,, e^{''}_R)^{\dagger}$, the charged-fermion mixing matrix is given by
\begin{align}
\label{eq:charge-fermion-mixing-matrix}
m_{e^{'}}=\frac{1}{\sqrt{2}}\left(
\begin{array}{cc}
  h_f\,v &  h_{c,\Phi }\,v_{\Phi } \\
  h_{b,\Phi }\,v_{\Phi } &  h_h\,v \\
\end{array}
\right)\,.
\end{align}
This matrix can be diagonalized by the biunitary transformation
\begin{align}
V^Lm_{e^{'}}(U^R)^{\dagger}=m_{e'_n}^{\text{diag}}\,,
\end{align}
where the mass eigenstates $e'_n=(e'_L\,, (e'_R)^{\dagger})_n$ are defined by
\begin{align}
\label{eq:dark-charge-fermions}
e'_{Ln}=&V^L_{ni}E_{Li}=\begin{pmatrix}
\cos\theta_L & \sin\theta_L \\
-\sin\theta_L & \cos\theta_L 
\end{pmatrix}
\begin{pmatrix}
e'_L \\ e^{''}_L
\end{pmatrix}\,, \quad
(e'_{R})^{\dagger}_n=U^R_{ni}(E_{Ri})^{\dagger}=\begin{pmatrix}
\cos\theta_R & \sin\theta_R \\
-\sin\theta_R & \cos\theta_R 
\end{pmatrix}
\begin{pmatrix}
e_R^{'\dagger} \\ e^{''\dagger}_R
\end{pmatrix}\,,
\end{align}
where $\theta_{L,R}$ are the two mixing angles.

Also, in this model, there are two new neutral dark Dirac fermions
\begin{align}
\label{eq:dark-neutral-fermions}
    N=&\begin{pmatrix}
    N'_L\\
    N''_R
    \end{pmatrix}\,, &
    \chi=&\begin{pmatrix}
    \chi_L\\
    \chi_R
    \end{pmatrix}\,, 
\end{align}
such that $N$ will be a heavy fermion and $\chi$ will be the candidate for the DM particle. 

Finally, the mixing matrix for the neutral gauge sector is given by
\begin{align}
m_Z^2=\left(
\begin{array}{ccc}
 \frac{1}{4}  g_1^2 \,v^2& -\frac{1}{4}  g_1 g_2\,v^2 & 0 \\
 -\frac{1}{4} g_1 g_2 \,v^2& \frac{1}{4}  g_2^2\, v^2 & 0 \\
 0 & 0 & 9 g_B^2 \,v_{\Phi }^2 \\
\end{array}
\right)\,,
\end{align}
where $g_1, g_2, g_B$ are the $\operatorname{U}(1)_Y$, $\operatorname{SU}(2)_L$ and $\operatorname{U}(1)_B$ gauge couplings. After diagonalization, the mass eigenstates are given by three neutral gauge bosons $(\gamma, \,Z, \,Z')$, where $\gamma$ is the photon field, $Z$ is the SM gauge boson, and $Z'$ is a new gauge boson with a mass $m_{Z'}=3g_B v_{\Phi}$.

\subsection{Dirac neutrino masses}
\label{sec:neutrinos}
When the scalar fields $H$ and $\Phi$ acquire VEVs, neutrinos obtain Dirac masses via the five-dimensional effective operator in Eq.~\eqref{eq:nmo56}, whose one-loop realization is shown in Fig.~\ref{fig:zee}. The diagram yields the mass matrix
\begin{align}
\label{eq:neutrino-mij-matrix}
\mathcal{M}_{i\alpha}=&\sum_{\beta=1}^2
h_d^{i\beta}\times \Lambda_{\beta}\times h_e^{\alpha\beta}\,,
\end{align}
where $\Lambda_\beta$ is the loop factor given by
\begin{align}
\label{eq:Lambda}
\Lambda_{\beta} = & \dfrac{1}{16\pi^2}\sum_{j=1}^4 Z^+_{j,\alpha}\,Z^+_{j,\alpha+2}
\sum_{n=1}^{2} V^L_{n1}\,U^R_{n1}\,m_{e'_{n}} 
\times \left[\dfrac{m_{e'_{n}}^2 \ln(m_{e'_{n}}^2) - m_{H_j^+}^2 \ln(m_{H_j^+}^2 )}{\left(m_{e'_{n}}^2-m_{H_j^+}^2\right)}\right]\,,
\end{align}
where $Z^+_{i,j}$, $V^L_{ij}$ and $U^R_{ij}$ are the rotation matrices defined in Eqs.~\eqref{eq:charge-higgs-diagonalization} and \eqref{eq:dark-charge-fermions}. $m_{e'_n}$, $m_{H_j^+}$ are the masses of the dark electrons and the charged scalars that are rotating in the loop, respectively.

Neutrino oscillation data at $3\sigma$~\cite{deSalas:2017kay} allow us to set the values of the Yukawa couplings in the Eq.~\eqref{eq:lagrangian-U1B}. In the basis where $\nu_R^{\alpha}$ are mass eigenstates, the mass matrix~\eqref{eq:neutrino-mij-matrix} can be written as~\cite{Kanemura:2011jj}
\begin{equation}
\label{eq:PMNS-relation}
\mathcal{M}_{ik}=(U_{\text{PMNS}})_{ik}\,(m_{\nu})_{k}\,,
\end{equation}
where $U_{\text{PMNS}}$ is the Pontecorvo-Maki-Nakagawa-Sakata matrix~\cite{Maki:1962mu} and $(m_{\nu})_k$ are the neutrino mass eigenvalues. 
Comparing the Eqs.~\eqref{eq:neutrino-mij-matrix} and~\eqref{eq:PMNS-relation}, we have 10 unknown parameters, $h_e^{\alpha \beta}, h_d^{i\alpha }$, and 9 equations. We can further simplify our analysis by imposing normal ordering for neutrino masses, $m_{\nu 1}<m_{\nu 2}<m_{\nu 3}$\footnote{Alternatively, we could choose inverted
ordering for the masses, $m_{\nu 3}<m_{\nu 1}<m_{\nu 2}$, and the analysis would proceed the same way.}, and leave the couplings $h_e^{\alpha \beta}$ as free parameters. We obtain the following relations:
\begin{eqnarray}
\label{eq:he-and-hd}
h_e^{\alpha \beta}&=& \text{free}\,, \nonumber\\
h_d^{i 1}&=& -\dfrac{1}{\Lambda_1}
\left(\dfrac{h_e^{23} m_{\nu 2} U_{i 2} - h_e^{22} m_{\nu 3} U_{i 3}}{h_e^{13} h_e^{22}- h_e^{12} h_e^{23} }\right), \nonumber\\
h_d^{i 2}&=& +\dfrac{1}{\Lambda_2}
\left(\dfrac{h_e^{13} m_{\nu 2} U_{i 2} - h_e^{12} m_{\nu 3} U_{i 3}}{h_e^{13} h_e^{22}- h_e^{12} h_e^{23} }\right)\,.
\end{eqnarray}
By construction, these Yukawa couplings will reproduce the current neutrino oscillation data~\cite{deSalas:2017kay}.

This neutrino mechanism can be probed at the LHC by searching for the electroweak-scale set of vector-like iso-singlet-doublet fermions~\cite{Restrepo:2015ura,Longas:2015sxk,Restrepo:2019soi}, 
and could be distinguished from the similar Dark-Zee mechanism with Majorana neutrinos by the absence of 
the extra inert scalar doublet~\cite{Longas:2015sxk}.

\subsection{Dark matter}
\label{sec:DM}

In this work, the Dirac fermion $\chi\,=\, \left(\chi_L,\,\chi_R \right)$ is the DM candidate. The processes involved in the calculation of the DM relic abundance include $\chi\bar{\chi} \to l_i\bar{l_i}$, $\chi\bar{\chi} \to q_i\bar{q_i}$, $\chi\bar{\chi} \to \nu_i\bar{\nu_i}$, $\chi\bar{\chi} \to h_k h_k$, $\chi\bar{\chi} \to h_k Z$, $\chi\bar{\chi} \to h_k Z'$, $\chi\bar{\chi} \to Z'Z'$. 

The evolution of the relic abundance of DM follows the standard WIMP freeze-out mechanism~\cite{Kolb:1990vq}. In the early radiation-dominated era of the Universe, $\chi$ was in thermal equilibrium with the primordial plasma. As the Universe cooled down to a temperature below the DM mass, the DM annihilation rate was overtaken by the expansion of the Universe, $\Gamma \ll H$, and a relic density of DM was frozen-out. 
The current relic abundance of DM depends on the thermally-averaged annihilation cross-section $\langle \sigma v\rangle \, \approx\, a\,+\,b v^2\,+\,\mathcal{O}(v^4)$, which yields~\cite{Kolb:1990vq, Srednicki:1988ce}
\begin{equation}
\label{eq:RelicAbundance}
\Omega_{\chi} h^{2} =\frac{2.04 \times 10^{9} x_{f}}{M_{\mathrm{Pl}} \sqrt{g_{*}(m_\chi)}\left(a+3 b / x_{f}\right)}\,,
\end{equation}
where $g_*(m)$ is the effective number of degrees of freedom, $x_f\equiv m/T$ and $h$ is today's Hubble parameter in units of $100~{\rm km/s/Mpc}$. 
 
\begin{figure}[t]
\centering
\includegraphics[scale=0.5]{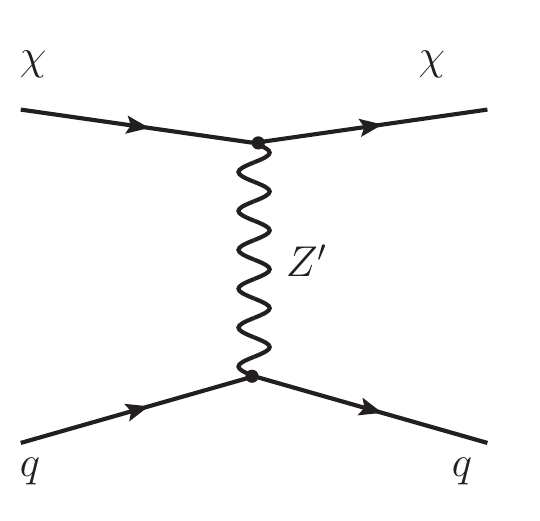}
\includegraphics[scale=0.5]{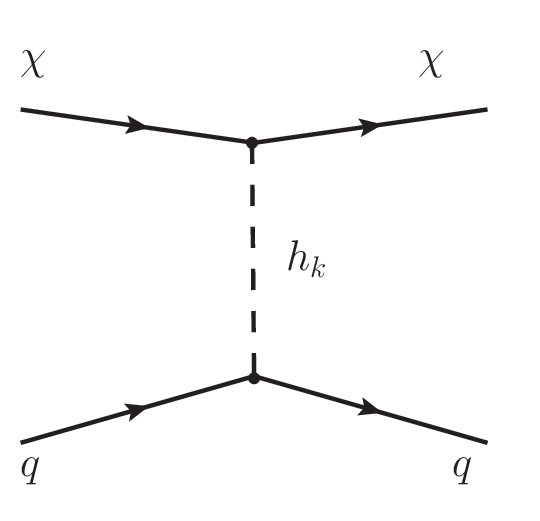}
\caption{SI independent DM-nucleon interactions: Vector(left) and scalar (right) portals.}
\label{fig:SI-diagram}
\end{figure}
This model allows for direct detection signatures since the DM scatters with nucleons through the $t$-channel with the $Z^{'}$ and scalars $h_k$ gauge boson as shown in Fig.~\ref{fig:SI-diagram}. The spin-independent (SI) cross-section for the interaction via the $Z^{'}$ gauge field is given by~\cite{Duerr:2014wra}
\begin{align}
\label{eq:SI}
\sigma^{\text{SI}}_{N}= \dfrac{1}{4\pi}\dfrac{M_N^2\,m_{\chi}^2}{(M_N\,+\,m_{\chi})^2}\dfrac{g_B^4}{M_{Z'}^4}\,,
\end{align}
where $M_N$ is the nucleon mass (neutron or proton). We only show here the vector interaction because it is the dominant process. This was verified numerically in Sec.~\ref{sec:scan}.

\subsubsection{Dark matter results}
\label{sec:scan}

\begin{figure}[t]
\begin{center}
\includegraphics[scale=0.6]{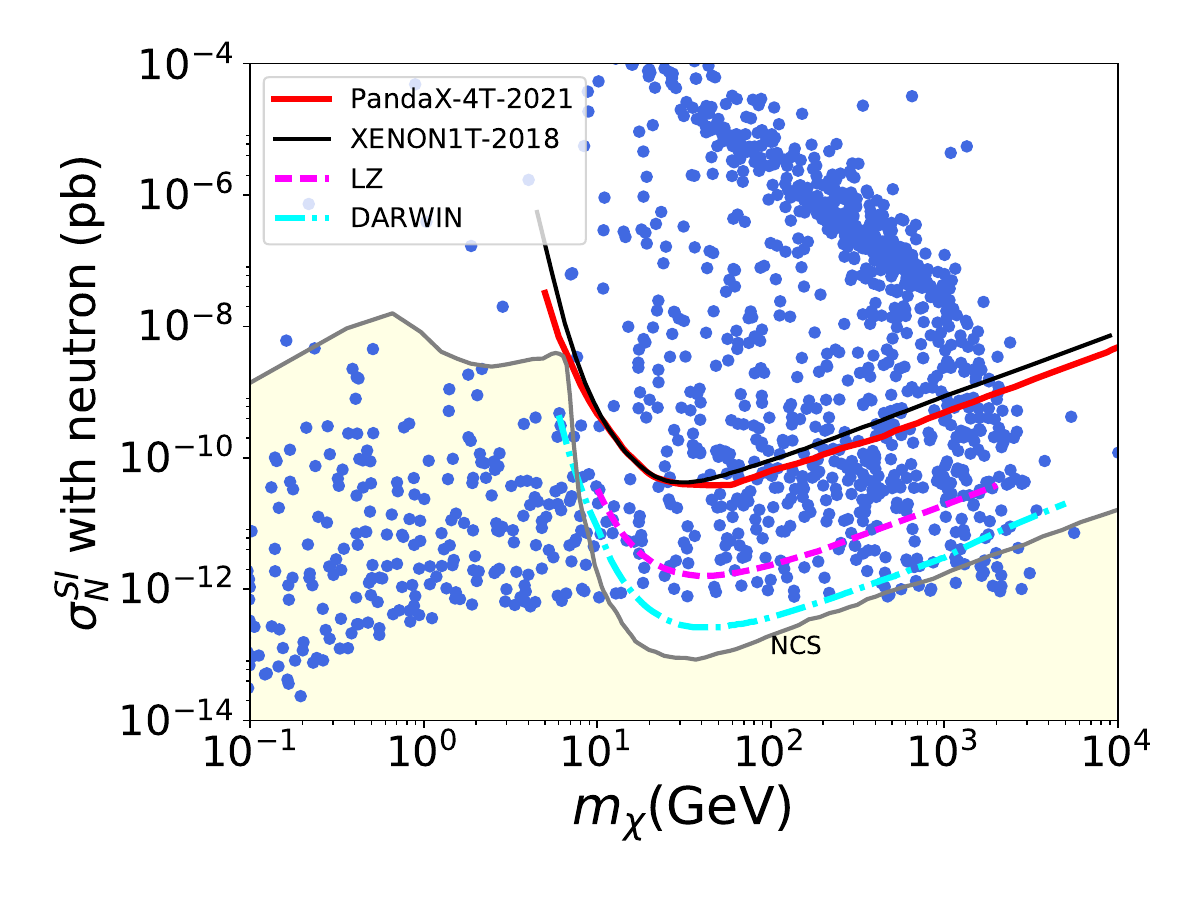}
\end{center}
\caption{The SI cross-section (blue dots) and the current experimental constraints from PandaX-4T~\cite{PandaX-4T:2021bab}, XENON1T~\cite{Aprile:2018dbl}, and prospects from LZ~\cite{Akerib:2018lyp} and DARWIN~\cite{Aalbers:2016jon}. 
We also show the Neutrino Coherent Scattering (NCS)~\cite{Cushman:2013zza, Billard:2013qya, Billard:2021uyg, OHare:2021utq}.}
\label{fig:SI}
\end{figure}

To study the phenomenology of this specific $\operatorname{U}(1)_B$ model, we have performed a random scan of the parameter space, varying the free parameters as described in Table~\ref{tab:scan}.
\begin{table}[t]
\centering
\begin{tabular}{c|c} 
\hline
Parameter & Range\\
\hline
$v_\Phi/$GeV & $10^{2}-2\times 10^{4}$ \\
$g_B$ & $10^{-6}-1$ \\
$h_{\{a,b,c\},\{S,\Phi\}}$, $h_g$, $h_h$ & $10^{-4}-1$ \\
$h_{e}$ & $10^{-5}-1$ \\
$\kappa_S^{\alpha\alpha}/$GeV, $\kappa_{\Phi}^{\alpha\alpha}/$GeV & $10^{-4}-1$ \\
$(\mu_S^2$, $\mu_{\sigma_{1}^\alpha}^2$, $\mu_{\sigma_{2}^\alpha}^2)/$GeV$^2$ & $10^{6}-10^{10}$\\
$\lambda_{\Phi}$ & $10^{-4}-1$ \\
$\lambda_{\Phi H}$, $\lambda_{S H}$, $\lambda_{{\sigma} H}$, $\lambda_{{\sigma'} H}$ & $10^{-5}-0.1$ \\
\hline
\end{tabular}
\caption{Scan ranges for free parameters of this model. }
\label{tab:scan}
\end{table}
We implemented the model in~\texttt{SARAH}~\cite{Staub:2008uz,Staub:2009bi,Staub:2010jh,Staub:2012pb,Staub:2013tta}, and coupled it to the \texttt{SPheno}~\cite{Porod:2003um,Porod:2011nf} routines to obtain the spectrum. 
To compute the DM relic density, we used~\texttt{MicrOMEGAs 5.0.4}~\cite{Belanger:2006is}, which includes the channels discussed above and some special processes such as coannihilations and resonances~\cite{Griest:1990kh}. 
Out all points in the parameter space, we selected the models that yield the current value for the DM relic density, $\Omega_{\chi} h^{2} =0.1200 \pm 0.0012$ to $3\,\sigma$~\cite{Aghanim:2018eyx} 
and reproduce the neutrino parameters, following the analysis described in Sec.~\ref{sec:neutrinos}.
For those models, we computed the SI DM-nucleus scattering cross-section~\eqref{eq:SI}, and checked it against the current experimental bounds from PandaX-4T~\cite{PandaX-4T:2021bab}, XENON1T~\cite{Aprile:2018dbl}, and prospective constraints from LZ~\cite{Akerib:2018lyp} and DARWIN~\cite{Aalbers:2016jon}. Additionally, we have filtered out those points that are inconsistent with monojet searches at the LHC~\cite{CMS:2017jdm}. The results are shown in Fig.~\ref{fig:SI}.

In Fig.~\ref{fig:SI}, the models under the red continuous line and $m_{\chi}\gtrsim 10$ (GeV) survive the current experimental limits imposed by PandaX-4T~\cite{PandaX-4T:2021bab} and could be explored in the near future by the LZ~\cite{Akerib:2018lyp} and DARWIN~\cite{Aalbers:2016jon} experiments.
It is noteworthy that there are points in the figure that fall into the neutrino floor (yellow region). They could be confused with neutrino coherent 
scattering (NCS) with nucleons, and they need a special analysis that is beyond the scope of this work.

An additional component of our analysis is the calculation of the spin-dependent (SD) WIMP-neutron cross-section using~\texttt{MicrOMEGAs}. We found that all models under the red line have a SD cross-section $\sigma^{\text{SD}} \le 10^{-43}$ \text{pb}, which falls below the experimental constraints from XENON1T~\cite{Aprile:2019dbj}. Furthermore, the SD cross-section obtained is below the prospects as LZ~\cite{Akerib:2018lyp} and DARWIN~\cite{Aalbers:2016jon}.

\section{Dark $CP$ violation and electroweak baryogenesis}
\label{sec:baryogenesis}
The third goal of this work is to generate the baryonic asymmetry in the Universe. The standard lore lists the conditions for a theory of baryogenesis as the {\it Sakharov conditions}~\cite{Sakharov:1967dj}: Violation of charge $C$, charge-parity $CP$, and there must exist processes that occur after exiting thermal equilibrium and violate baryon number, $B$. To satisfy these conditions, we will adapt the mechanism presented in \cite{Carena:2018cjh,Carena:2019xrr} in which $CP$ violation occurs in the dark sector and is mediated to the SM sector by the new $Z'$ gauge boson. Baryogenesis results from the dynamics of the same hidden-sector fields that are also responsible for dark matter and neutrino masses. The goal of this section, rather than presenting a new mechanism for baryogenesis, is to show that models of scotogenic Dirac neutrino masses and dark matter can easily accommodate electroweak baryogenesis in the same fashion as \cite{Carena:2018cjh,Carena:2019xrr}\footnote{Ref.~\cite{Cline:2017qpe} presented another mechanism where $CP$ violation occurs in the hidden sector and there is a Yukawa coupling between the dark and SM fermions.}.

In the scalar sector of the model, the key fields in the mechanism are the scalars $H = h/\sqrt{2}$ and $s = |S|$. A strong first-order phase transition is incorporated into this scenario through the evolution of the VEVs of these two fields, whose potential can be rewritten as~\cite{Cline:2017qpe}\footnote{We assume that the field $\Phi$, which has a much larger VEV is integrated out and does not play any role in the baryogenesis mechanism beyond providing mass terms in the Lagrangian.} 
\begin{align}
    V(h,s)\,=\,\frac{\lambda_H}{4}\left(h^2- v^2\right)^2 +\frac{\lambda_S}{4}\left(s^2-w^2\right)^2 + \frac{\lambda_{SH}}{2}h^2 s^2\,,
\end{align} 
where $v$ and $w$ are the VEVs for $h$ and $s$, respectively,  at the minimum of the potential. We require two stable minima, $(0,\,w_0)$ and $(v,\,w)$, for this potential and a tree-level barrier between them. Furthermore, the global minimum at zero temperature must be at $v_\text{EW}\equiv v(0) = 246\, {\rm GeV}$ and $w(0)=0$. Following the potential analysis in \cite{Espinosa:2011ax}, we find that these conditions can be satisfied if
\begin{align} \label{eq:lamS_cond1}
    \lambda_{SH}>0, \quad \lambda_H \lambda_S -\frac{1}{4} \lambda_{SH}^2 < - \frac{\lambda_{SH} m_s^2}{2 v^2}\,.
\end{align} 

At high temperatures, $\Phi$ breaks the $\operatorname{U}(1)_B$ symmetry and the global minimum of the potential is given by $\left(0, w_0(T) \right) $ and the electroweak symmetry is exact. As the temperature decreases, the electroweak minimum forms with $ (v(T), w(T))$. At the critical temperature $T_c$, both minima are degenerate. For lower temperatures, $T<T_c$, the electroweak minimum becomes the global minimum. The finite-temperature effective potential is given by 
 \begin{align}
    V_T(h,s)\,=\,& \frac{\lambda_H v_c^4}{4}\left(\frac{h^2}{v_c^2} +\frac{s^2}{w_c^2} -1\right)^2 + \frac{\lambda_H v_c^2}{m_{s,c}^2 w_{0,c}^4}h^2 s^2 
     + (T^2-T_c^2)(c_h h^2 +c_s s^2)\,, 
\end{align}
where the subscript $c$ denotes the quantity at $T=T_c$. The coefficients $c_s$ and $c_h$ correspond to one-loop thermal corrections and are given by 
\begin{align}
    c_h\,=\,&\frac{1}{48} \left(9g_2^2+ 3g_1^2 +12 y_t^2 +24 \lambda_H +\lambda_{SH}\right)\, ,\quad
    c_s\,=\,  \frac{1}{12} \left (3 \lambda_S + 2 \lambda_{HS} \right)\,. 
\end{align} 
An additional condition, to ensure that the global minimum for this potential is the broken one when $T=0$, is 
\begin{equation} \label{eq:lamS_cond2}
    \frac{c_h}{c_s} > \sqrt{\frac{\lambda_H}{\lambda_S}}.
\end{equation}

\begin{figure}[t]
\centering
\includegraphics[scale=0.4]{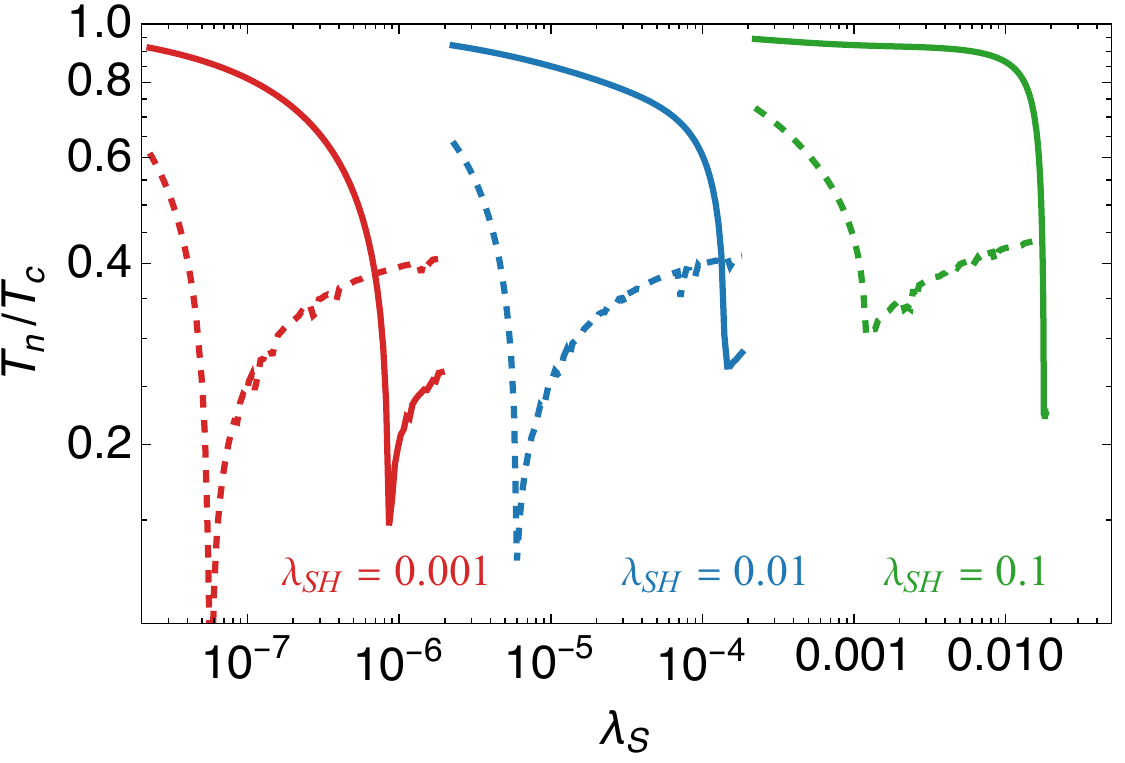}
\includegraphics[scale=0.41]{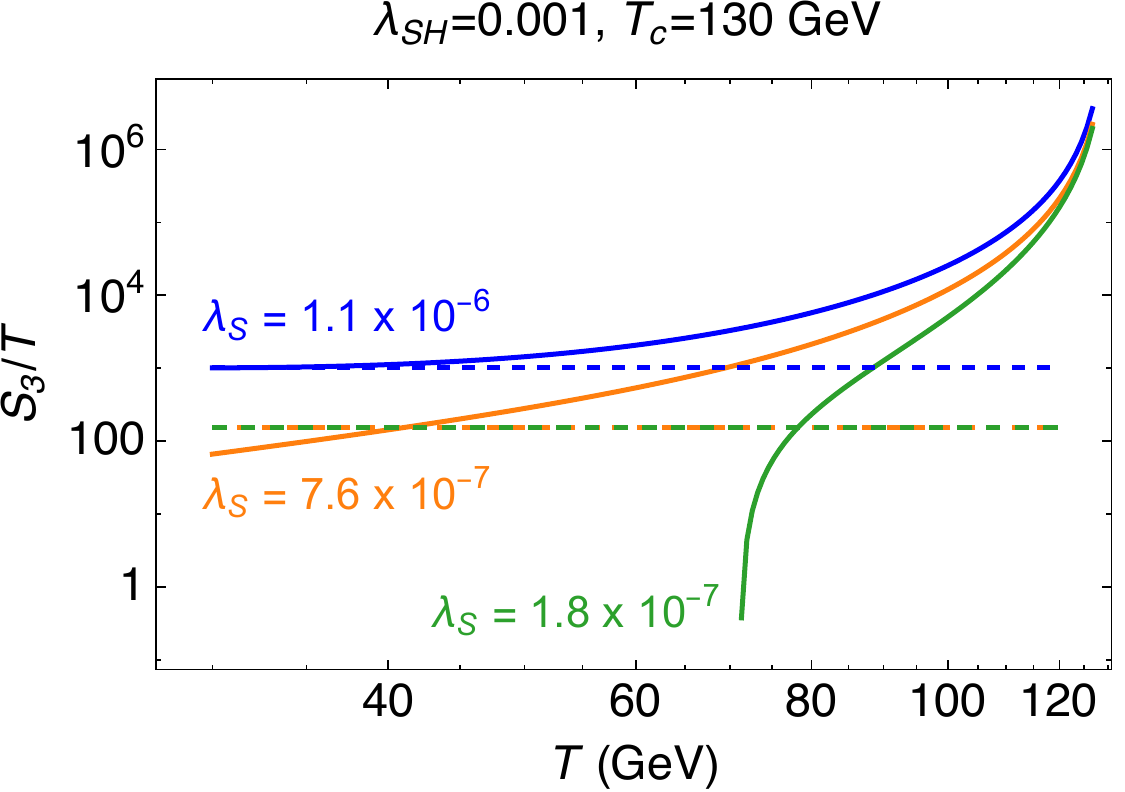}
\caption{{\it Left:} $T_n$ dependence on $\lambda_S$ for different choices of $\lambda_{SH}$. Solid lines correspond to $T_c=130 {\,\rm GeV}$ and dashed lines to $T_c=110\,{\rm GeV}$. The range of the lines is bounded by Eqs.~\eqref{eq:lamS_cond1} and \eqref{eq:lamS_cond2}. {\it Right:} The solid lines show the $T$-dependence of $S_3(T)/T$ in the thin-wall approximation for different values of $\lambda_S$ while keeping $\lambda_{SH}$ fixed. The dashed lines show the value of $S_3(T_n)/T_n$ for each of the choices of $\lambda_S$. Notice that the green and orange dashed lines overlap. }
\label{fig:Tn_plots}
\end{figure}

Using the thin-wall approximation~\cite{Coleman:1977py}, the nucleation temperature, $T_n$, is defined by the condition~\cite{Cline:2017qpe}
\begin{equation} \label{eq:S3}
    \exp \left (-S_3 /T_n \right) \,=\, \frac{3}{4\pi}\left(\frac{H(T_n)}{T_n} \right)^4 \left(\frac{2\pi\,T_n}{S_3}\right)^{\frac{3}{2}} \,,
\end{equation} where $S_3$ is the Euclidean action of the bubble and $H(T)$ is the Hubble rate. In this approximation, to describe the bubble wall profile, we use the ansatz in which the space dependence of the fields is given by 
\begin{equation} \label{eq:ansatz}
    h(z) \,=\, \frac{1}{2}v(T_n) \left(1- \tanh \left(z/L_w \right) \right)\,,\qquad
    s(z) \,=\, \frac{1}{2}s_0 \left(1+ \tanh \left(z/L_w \right) \right)\,, 
\end{equation} 
where $z$ is the direction normal to the wall and $L_w$ is the wall width and $s_0 \equiv w_0(T_n)$. At the first-order phase transition, bubbles nucleate and expand through the primordial plasma, causing perturbations on the particle and antiparticle densities. For a given set of parameters and critical temperature $T_c$, we obtain $T_n$ by solving numerically Eq.~\eqref{eq:S3}. In Fig.~\ref{fig:Tn_plots}, we show the dependence of $T_n$ on the coupling $\lambda_S$ for different values of $\lambda_{SH}$ and two choices of $T_c$. As we will explain below, in our model, the nucleation temperature does not turn out much smaller than the critical temperature. Hence, the relevant regions in these curves are those on the lower values of $\lambda_S$, to the right of the inverted peak.

The velocity and width of the wall can be calculated following the algorithm presented in \cite{Konstandin:2014zta,Friedlander:2020tnq,Lewicki:2021pgr}, where the ansatz \eqref{eq:ansatz} is not used. Instead, starting from initial guesses for $v_w$ and $L_w$, small iterative variations are made until the equations of motion for the bubble profile are successfully solved. This is a computationally and numerically costly process. However, in the analysis presented in \cite{Friedlander:2020tnq}, the calculation of the wall velocity for a scalar potential like ours showed that there is a correlation between the wall velocity and the supercooling parameter $r\equiv v(T_n)T_c/v_cT_n$. As long as $r$ remains relatively close to the unity, $r\lesssim 1.15$, the models lead to strong phase transitions with subsonic velocities, $v_w<1/\sqrt{3}$. In this work, we will adopt this condition and vary the wall velocity in the range $0.1 \lesssim v_w \lesssim 1/\sqrt{3}$. On the other hand, the wall thickness can be approximated by \cite{Espinosa:2011eu,Cline:2017qpe}
\begin{equation}
L_w \simeq \left(\frac{2.7 (v_c^2+w_c^2)}{v_c^2(\lambda_{SH}w_c^2-2\lambda_Hv_c^2)}\left(1+\frac{\lambda_{SH}w_c^2-2\lambda_Hv_c^2}{4\lambda_H v_c^2} \right) \right)^{1/2}\ .
\end{equation}

$P$ and $CP$ violation is incorporated by adding a term \footnote{In the same spirit of \cite{Carena:2019xrr}, we could add other terms with different powers of $S,\,S^*$ or $\Phi$, however, we keep only this one for simplicity. We also ignore the backreaction of $\delta V$ into the spontaneous breaking of $U(1)_B$.}
\begin{equation}
    \delta V(S,\Phi) \,=\, \tilde{\lambda}_{\Phi S} \Phi^{*2} S^2\,,
\end{equation} which is invariant under $\operatorname{U}(1)_B$ but generates a term $\delta V \sim \langle \Phi \rangle^2 S^2$ after $\Phi$ acquires a VEV. This term in the potential is minimized by setting the phase of $S$ equal to $\pi/2$. The chiral fields, $\chi_L$ and $\chi_R$, couple to both scalar fields $\Phi$ and $S$. After the spontaneous breaking of $\operatorname{U}(1)_B$, those chiral fields acquire an effective mass 
\begin{equation}
    M_\chi \,=\ m_{\chi} +\lambda e^{i \theta} s\,,
\end{equation} 
where $m_{\chi} = h_{a,\Phi} \langle \Phi \rangle$ and we have parametrized the Yukawa coupling $h_{a,S}\,=\, \lambda e^{i\theta-i\pi/2}$. The parameters $m_{\chi}$ and $\tilde{\lambda}_{\Phi S}$ are taken to be real, which is possible due to the freedom to redefine the fields. However, we cannot eliminate the complex phase in the second term. This phase violates $CP$ and will lead to opposite signs in the perturbations of particles and antiparticles, resulting in a net asymmetry in the interior of the bubble, which is not washed out if the condition $v(T_n)/T_n > 1$ is satisfied~\cite{Cline:2017qpe}.

The evolution of the particle and anti-particle distribution functions is obtained from the Boltzmann equations, which are recast as the diffusion equation for the re-scaled chemical potential, $\xi_i(z)\,\equiv\,{\mu_i(z)}/{T} \,=\,{6}\left(n_i - \overline{n}_i \right)/{T^3}$,
\begin{equation}
- D_L \xi''_{\chi_L} -v_w\xi'_{\chi_L} +\Gamma_L(\xi_{\chi_L}-\xi_{\chi_R})\,=\,S_{\cancel{CP}}\,,
\end{equation} where $D_L$ is the diffusion constant for $\chi_L$, which is related to the scattering rate $\Gamma_L$ by $D_L=\langle v_{p_z}^2\rangle/3\Gamma_L$. Here, $\langle\,\rangle$ means thermal average and $z$ denotes the normal direction to the wall. $S_{\cancel{CP}}$ is $CP$-violating source that results from the variation of $\theta$~\cite{Cline:2000nw},
\begin{eqnarray} \label{eq:diffeq}
S_{\cancel{CP}}\,=\,-\frac{\lambda}{2}\frac{v_w D_L}{\langle v_{p_z}^2  \rangle T} \left\langle \frac{|p_z|}{\omega^2}  \right\rangle \left(M_\chi^2\, \theta' \right)''\,,
\end{eqnarray} 
where $v_w$ is the wall's velocity and 
\begin{align}
& \quad \langle v_{p_z}^2  \rangle \,=\, \frac{3x+2}{x^2 +3x+2}\,, \quad \left\langle \frac{|p_z|}{\omega^2}  \right\rangle \,=\,\frac{(1-x)e^{-x} + x^2 E_1(x) }{4m_\chi^2 K_2(x)}\, ,\quad x\equiv m_\chi /T\,,  \nonumber \\
&  \left(M_\chi^2 \theta' \right)''\,=\, \frac{m_\chi s_0 \lambda \left(-2+\cosh \left(\frac{2z}{L_w} \right) \right)\sin \theta}{L_w^3 \cosh^4 \left(\frac{z}{L_w} \right)}\, .
\end{align} $E_1(x)$ is the error function and $K_2(x)$ is the modified Bessel function of the second kind.

The novelty of this mechanism, as presented by the authors of~\cite{Carena:2018cjh,Carena:2019xrr}, is that the chiral particles, for which the asymmetry is initially generated, do not couple to the $\operatorname{SU}(2)_L$ current, but instead give rise to a non-zero $\operatorname{U}(1)_B$ charge density in the proximity of the wall. This results in a $Z'$ background that couples to the SM fields with $\operatorname{U}(1)_B$ charge,
\begin{equation}
    \langle Z'_0 \rangle \,=\, \frac{ g_B\,(q_{\chi_L}-q_{\chi_R})T_n^3}{6 M_{Z'}} \int_{-\infty}^\infty \operatorname{d}z'\, \xi_{\chi_L}(z')\, e^{-M_{Z'}|z-z'|}\, , 
\end{equation} where $\xi_{\chi_L}$ is given by the solution to Eq.~\eqref{eq:diffeq}, which is given by~\cite{Carena:2018cjh,Carena:2019xrr}
\begin{equation}
\xi_{\chi_L} (z)\,=\, \int_{-\infty}^{\infty} \operatorname{d}z_1 G(z-z_1)\,S_{\cancel{CP}},
\end{equation}
where $G(z)$ is thee Green's function
\begin{equation}
G(z)\,=\,\frac{D_L^{-1}}{k_+-k_-}\left\{ \begin{array}{c}
     e^{-k_+ z},\,\, z\geq 0  \\
     e^{-k_- z},\,\, z< 0 
\end{array} \right. \,,\quad k_{\pm}\,\equiv\,\frac{v_w}{2D_L} \left(1\pm\sqrt{1+\frac{8\Gamma_L D_L}{v_w^2}} \right).
\end{equation}
The $Z'$ background generates a chemical potential for the SM quarks
\footnote{At $T=T_c$, the heavy fields $L'_L$ and $L''_R$ are already thermally decoupled and the $\operatorname{U}(1)_B$ current in the plasma is anomalous with respect to $\operatorname{SU}(2)_L$. This allows the generation of the non-zero chemical potential~\cite{Carena:2019xrr}.}, 
\begin{equation}
\mu_{Q}(z)\,=\, \mu_{d_R,u_R}(z)\,=\, 3 \times \frac{1}{3} \times g_B \langle Z'_0 (z) \rangle, 
\end{equation} 
which sources a thermal-equilibrium asymmetry in the quarks~\cite{Carena:2019xrr}, $ \Delta n_{Q}^{\rm EQ} (z) \,\sim \, T_n^2 \, \mu_{Q}(z)$. 

Finally, the baryon-number asymmetry is then given by 
\begin{equation}
\label{eq:nB}
  n_B\,=\,  \frac{\Gamma_{\rm sph}}{v_w} \int_0^{\infty}\operatorname{d}z\, n_{Q}^{\rm EQ} (z)\, \exp\left(-\frac{\Gamma_{\rm sph}}{v_w}\,z\right)\,,
\end{equation} 
where $\Gamma_{\rm sph}$ is the sphaleron rate, $\Gamma_{\rm sph} \simeq 120\alpha_W^2T_n$. The baryon-to-photon-number ratio is then obtained by
\begin{equation}
 \eta_B \,=\, \frac{n_B}{s(T_n)},\quad s (T) \,\equiv\,\frac{2\pi^2}{45} g_{*S}(T)\, T^3\,, 
\end{equation} where $g_{*S}(T)$ is the effective number of relativistic degrees of freedom.

Our goal is to find what regions of the parameter space yield a baryon-number asymmetry approximately equal to the measured value~\cite{Zyla:2020zbs},
\begin{align}
\label{eq:eta-value}
0.82\times 10^{-10} < \eta_B < 0.92\times 10^{-10}\,.
\end{align}

\begin{figure}[t]
\centering
\includegraphics[scale=0.6]{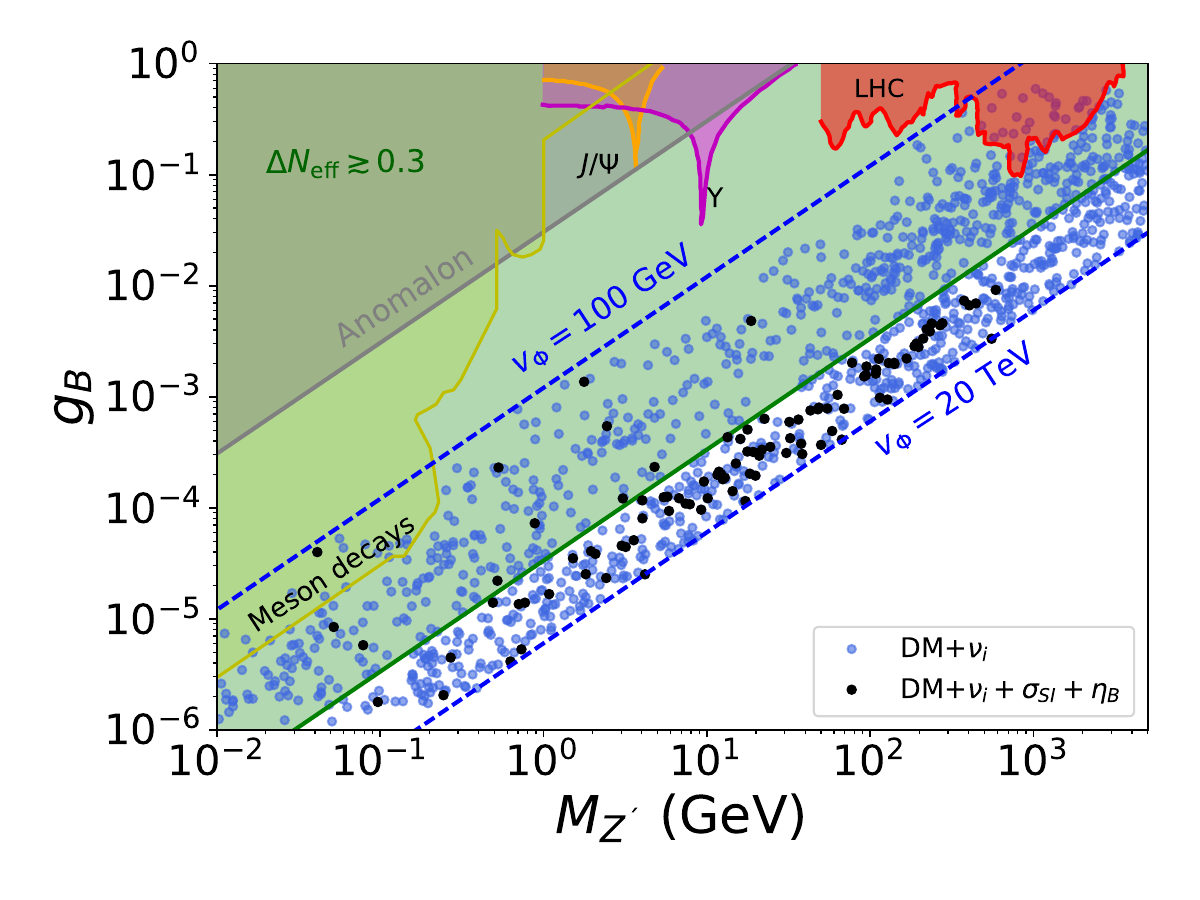}
\caption{The blue points are the models shown in Fig.~\ref{fig:SI} that fulfill the relic abundance of DM and the neutrino masses. The black points are those that are not excluded by direct detection of DM and give the observed baryon asymmetry at the Universe. The green-shaded region is in tension with the measured number of relativistic degrees of freedom~\cite{PhysRevD.101.035004}. The other shaded areas show the exclusions coming from meson decays~\cite{Dror:2017nsg} (yellow), the hadronic widths of $\Upsilon$ and $J/\psi$ (purple and orange), LHC dijet searches~\cite{FileviezPerez:2018jmr} (red), and from non-detection of the anomaly-cancelling fermions~\cite{Dobrescu:2014fca} (gray).
}
\label{fig:gB-vs-MZp}
\end{figure}
We computed the baryon asymmetry for the $U(1)_B$ model presented in Sec.~\ref{sec:U1B-model} for those points in the parameter space that are consistent with neutrino masses and that produce the expected relic abundance of dark matter. For the free parameters that are not involved in neutrino mass and dark matter calculations, we performed a random scan in the ranges shown in Table~\ref{tab:scan-baryogenesis}. 

The results are displayed in the plane $M_{Z'}-g_B$ in Fig.~\ref{fig:gB-vs-MZp}. 
The blue points are models that fulfill the relic abundance of DM and neutrino masses (displayed also in Fig.~\ref{fig:SI}). The black points are the models that, in addition, yield the observed baryon asymmetry~\eqref{eq:eta-value} and are not excluded by direct detection of DM.
\begin{table}[t]
\centering
\begin{tabular}{c|c} 
\hline
Parameter & Range\\
\hline
$\theta$ & $(-\pi/2 \,, \pi/2)$  \\
$s_0/{\rm GeV} $ &  $(100,\,500)$\\
$T_c/{\rm GeV}$ &  $(100,\,150)$\\
$\lambda_{SH}$ &  $(10^{-5},\,0.1)$\\
$\lambda$ &  $(10^{-5},\,1.0)$\\
$v_w$ &  $(0.1, \, 0.5)$\\
\hline
\end{tabular}
\caption{Scan ranges for the free parameters that are involved in the baryogenesis mechanism. }
\label{tab:scan-baryogenesis}
\end{table}
We found that, for $M_{Z'}\gtrsim 800$ GeV, $\eta_B$ is always lower than the experimental value. This behavior is understood from the fact that the baryon asymmetry scales as $\sim g_B/M_{Z'}$. For illustrative purposes, we include in Table~\ref{tab:scan-average} the average values of the parameters that yield the black dots in Fig.~\ref{fig:gB-vs-MZp}.

\begin{table}[t]
\centering
\resizebox{\textwidth}{!}{
\begin{tabular}{c|c|c|c|c|c|c|c|c|c|c|c|c} 
\hline

\hline
&$M_{Z'}/{\rm GeV}$&$\log_{10}g_B$&$m_\chi/{\rm GeV}$&$\theta$&$\log_{10}\lambda_{SH}$&$\log_{10}\lambda_S$&$\log_{10}\lambda$&$s_0/{\rm GeV}$&$L_w\,T_n $ &$T_n/{\rm GeV}$&$T_c/{\rm GeV}$&$v_w$ \\
\hline
Average &86.49 & -3.1 &272.0 & 0.385 & -1.50 &-2.38& -0.64 &346.5& 4.48 &111.6& 131.2 &0.252\\
\hline
Std. dev. &80.0 & 0.94 &184.2 & 1.274 & 0.79 &1.20 &0.38 &93.6 &3.17 & 10.3& 6.1 &0.13\\
\hline
\end{tabular}
}
\caption{Average and standard deviation of the parameter values that yield the right baryon asymmetry.}
\label{tab:scan-average}
\end{table}

Models with low $v_\Phi$ and $m_\chi \gtrsim 10$ GeV are excluded by direct detection of DM. This is understood because according to Eq.~\eqref{eq:SI}, the SI cross-section scales as $1/v_\Phi^4$ which needs high values for $v_\Phi$ to past the current experimental bounds of PandaX-4T. In addition, in Fig.~\ref{fig:gB-vs-MZp}, we show the limits from the hadronic widths of $\Upsilon$ and $J/\psi$, constraints from the LHC dijet searches~\cite{FileviezPerez:2018jmr}, and constraints from meson decays~\cite{Dror:2017nsg}. Finally, the right-handed neutrinos contribute to the number of relativistic degrees of freedom. We computed this contribution~\cite{PhysRevD.101.035004} and required that $\Delta N_{\text{eff}}<0.3$, as measured by the Planck collaboration~\cite{Aghanim:2018eyx}. This results in $M_{Z'}/g_B \gtrsim 30$ TeV, which implies $v_{\Phi} \gtrsim 10$ TeV. This constraint imposes a strong restriction over our model as it excludes, in Fig.~\ref{fig:gB-vs-MZp}, the green region. The bottom line is that only the black points under the green continuous line $M_{Z'}/g_B = 30$ TeV reproduce the relic abundance of DM, neutrino masses, and the baryon asymmetry at the Universe while being allowed by direct detection of DM and cosmological constraints on $\Delta N_{\text{eff}}$.

\section{Conclusions} 
\label{sec:con}
We have added an explanation to the smallness of Dirac-neutrino masses in the framework of electroweak baryogenesis through the $Z'$ boson of a gauged baryon or lepton number, by using the required non SM-singlet fermions fields as the scotogenic particles of the Dirac-dark Zee model. In addition to generating Dirac masses for neutrinos and the expected relic density of dark matter, the dark sector is responsible for providing the necessary amount of $CP$ violation and the non-zero $Z'$ background to generate the baryon asymmetry in the Universe. We scanned the parameter space
requiring that these goals be achieved while being
compatible with direct detection experiments and big bang nucleosynthesis. We found that models with $800\,{\rm GeV}\,\gtrsim \,M_Z'\,\gtrsim\, {\rm 30\, {\rm MeV}}$ can satisfy all conditions with dark matter masses in the WIMP range.

\section*{Acknowledgments}
We want to thank Yue Zhang for very valuable feedback and illustrative conversations in the course of this work. The  work  of  DR  is  supported  by  Sostenibilidad  UdeA,  and the  UdeA/CODI  Grant~2020-33177, and FAPESP funding Grant 2021/11383-4. W.T. is supported by the National Science Foundation under Grant No. PHY-2013052.

\bibliographystyle{apsrev4-1long}
\bibliography{biblio}

\end{document}